\documentclass[english,12pt]{article}
          \usepackage{babel}
          \usepackage[T1]{fontenc}
          \usepackage[latin2]{inputenc}
          \usepackage{pslatex}
\def\th{\thinspace}

\begin{document}

\title{\bf Nonradial Modes in Classical Cepheids}
    \author{P. Moskalik$^1$ ~and~ Z. Ko{\l}aczkowski$^2$}
\date{$^1$ Copernicus Astronomical Center, Warszawa, Poland\break
      $^2$ Universidad de Concepcion, Concepcion, Chile ~and~
           Wroc{\l}aw University Observatory, Wroc{\l}aw, Poland}

\maketitle

\begin{abstract}

Systematic search for multiperiodicity in the LMC Cepheids
(Moskalik, Ko{\l}aczkowski \& Mizerski 2004) has led to discovery
of low amplitude nonradial modes in a substantial fraction of
overtone pulsators. We present detailed discussion of this new
type of multimode Cepheid pulsators and compare them to similar
nonradial pulsators discovered among RR~Lyrae stars. Finally, we
show first detections of secondary nonradial modes in FU/FO
double-mode Cepheids.

\end{abstract}

\section*{LMC Cepheids: Data and Analysis}

\hskip 0pt Our search for multiperiodic variations in the LMC
Cepheids was performed with I-band DIA-reduced OGLE-II photometry
(\.Zebru\'n et~al. 2001). It spans 1000-1200\th days, with 250-500
flux measurement per star. The data was analyzed with a standard
consecutive prewhitening technique. First, we fitted the data with
a Fourier sum representing variations with the dominant frequency:

\begin{equation}
I(t) = \langle I\rangle + \sum_{k} {\rm A}_{k} \sin (2\pi k{\rm f} t + \phi_{k}).
\end{equation}

\noindent The frequency of the mode, ${\rm f}$, was also
optimized. For double mode Cepheids, a double frequency Fourier
sum was fitted. The residuals of the fit were then searched for
secondary frequencies. This was done with the Fourier transform,
calculated over the range of $0-5$\th c/d. In the next step, a new
Fourier fit with all frequencies identified so far was performed
and the fit residuals were searched for additional frequencies
again. The process was stopped when no new frequencies were
detected.

\smallskip

\hskip 0pt We have analyzed all fundamental (FU), first overtone
(FO) and double mode (FU/FO and FO/SO) Cepheids listed in the
OGLE-II catalogs (Udalski et~al. 1999; Soszy\'nski et~al. 2000),
nearly 1300 stars in total. Results for the FO/SO double mode
Cepheids have been presented elsewhere (Moskalik, Ko{\l}aczkowski
\& Mizerski 2006; Moskalik \& Ko{\l}aczkowski 2008). Here we
discuss in details our findings for single mode and for FU/FO
double mode pulsators.

\section*{First Overtone Cepheids}

\hskip 0pt The OGLE-II catalog list 462 first overtone (FO)
Cepheids. We have detected residual power in 64 of them. In 42
variables, which constitute 9\% of the entire LMC sample, we were
able to resolve this power into individual frequencies. We
consider two frequencies to be resolved if $1/\Delta{\rm f} <
600$\th days). For one object frequency resolution was achieved
with the MACHO data (Allsman \& Axelrod 2001), which is more than
twice longer. The complete inventory of resolved FO Cepheids is
presented in Table\th 1. Following notation originally
introduced for RR~Lyrae variables (Alcock et~al. 2000), we call
these stars FO-$\nu$ Cepheids. Consecutive columns of Table\th 1
give OGLE number of the star, primary and secondary periods ${\rm
P}_{1}$ and ${\rm P}_{\nu}$, frequency difference $\Delta{\rm f} =
{\rm f}_{\nu} - {\rm f}_{1}$, period ratio ${\rm P}_{\nu}/{\rm
P}_{1}$ and amplitude ratio ${\rm A}_{\nu}/{\rm A}_{1}$. All
parameters are determined from the least square fits to the data.

\medskip\bigskip

\begin{center}
{\bf Table 1:} FO-${\nu}$ Cepheids in OGLE-II LMC Sample
\end{center}

\begin{center}
\normalsize
\small
\begin{tabular}{lccccc}
\hline
\noalign{\smallskip}
~~OGLE ID         & $P_1$\th [day]
                                 & $P_{\nu}$\th [day]
                                                & $\Delta f$\th [c/d]
                                                                & $P_{\nu}/P_1$
                                                                              & $A_{\nu}/A_1$
                                                                                          \\
\noalign{\smallskip}
\hline
\noalign{\smallskip}
 ~~SC1--201683~~  & ~~2.319211~~ & ~~2.340033~~ & ~~--0.00384~~ & ~~1.00898~~ & ~~0.048~~ \\
                  &              & ~~3.069683~~ & ~~--0.10542~~ & ~~1.32359~~ & ~~0.022~~ \\
\noalign{\smallskip}
 ~~SC1--229541~~  & ~~1.216360~~ & ~~1.359217~~ & ~~--0.08641~~ & ~~1.11745~~ & ~~0.043~~ \\
\noalign{\smallskip}
 ~~SC2--158672~~  & ~~2.313334~~ & ~~1.477135~~ & ~~~~0.24471~~ & ~~0.63853~~ & ~~0.032~~ \\
\noalign{\smallskip}
 ~~SC2--208897~~  & ~~2.417847~~ & ~~2.346886~~ & ~~~~0.01251~~ & ~~0.97065~~ & ~~0.074~~ \\
\noalign{\smallskip}
 ~~SC2--283723~~  & ~~1.308907~~ & ~~1.425531~~ & ~~--0.06250~~ & ~~1.08910~~ & ~~0.070~~ \\
\noalign{\smallskip}
 ~~SC3--274410~~  & ~~2.446161~~ & ~~2.981644~~ & ~~--0.07342~~ & ~~1.21891~~ & ~~0.043~~ \\
                  &              & ~~3.099460~~ & ~~--0.08617~~ & ~~1.26707~~ & ~~0.033~~ \\
\noalign{\smallskip}
 ~~SC3--421512~~  & ~~3.186057~~ & ~~4.652311~~ & ~~--0.09892~~ & ~~1.46020~~ & ~~0.055~~ \\
\noalign{\smallskip}
 ~~SC4--36200~~   & ~~3.476962~~ & ~~3.912773~~ & ~~--0.03203~~ & ~~1.12534~~ & ~~0.054~~ \\
                  &              & ~~4.057330~~ & ~~--0.04114~~ & ~~1.16692~~ & ~~0.038~~ \\
\noalign{\smallskip}
 ~~SC4--131738~~  & ~~3.122245~~ & ~~3.780949~~ & ~~--0.05580~~ & ~~1.21097~~ & ~~0.032~~ \\
\noalign{\smallskip}
 ~~SC4--295932~~  & ~~4.166232~~ & ~~2.633729~~ & ~~~~0.13967~~ & ~~0.63216~~ & ~~0.023~~ \\
\noalign{\smallskip}
 ~~SC5--75989~~   & ~~2.238255~~ & ~~2.949657~~ & ~~--0.10775~~ & ~~1.31784~~ & ~~0.091~~ \\
\noalign{\smallskip}
 ~~SC5--138031~~  & ~~2.681369~~ & ~~3.249411~~ & ~~--0.06520~~ & ~~1.21185~~ & ~~0.029~~ \\
\noalign{\smallskip}
 ~~SC6--135695~~  & ~~1.864922~~ & ~~1.949614~~ & ~~--0.02329~~ & ~~1.04541~~ & ~~0.037~~ \\
\noalign{\smallskip}
 ~~SC6--135716~~  & ~~2.838799~~ & ~~2.780248~~ & ~~~~0.00742~~ & ~~0.97938~~ & ~~0.064~~ \\
\noalign{\smallskip}
 ~~SC6--267289~~  & ~~1.845746~~ & ~~1.906899~~ & ~~--0.01738~~ & ~~1.03313~~ & ~~0.039~~ \\
\noalign{\smallskip}
 ~~SC6--363194~~  & ~~2.797085~~ & ~~3.727012~~ & ~~--0.08920~~ & ~~1.33246~~ & ~~0.043~~ \\
\noalign{\smallskip}
 ~~SC7--344559~~  & ~~2.062992~~ & ~~2.140660~~ & ~~--0.01759~~ & ~~1.03765~~ & ~~0.068~~ \\
\noalign{\smallskip}
 ~~SC8--205108~~  & ~~3.515988~~ & ~~3.905659~~ & ~~--0.02884~~ & ~~1.11083~~ & ~~0.026~~ \\
\noalign{\smallskip}
 ~~SC8--224964~~  & ~~2.866473~~ & ~~3.189296~~ & ~~--0.03531~~ & ~~1.11262~~ & ~~0.041~~ \\
\noalign{\smallskip}
 ~~SC9--216934~~  & ~~4.001017~~ & ~~4.997407~~ & ~~--0.04983~~ & ~~1.24903~~ & ~~0.031~~ \\
\noalign{\smallskip}
 ~~SC9--230584~~  & ~~4.395271~~ & ~~5.672324~~ & ~~--0.05122~~ & ~~1.29055~~ & ~~0.059~~ \\
                  &              & ~~5.860579~~ & ~~--0.05689~~ & ~~1.33338~~ & ~~0.028~~ \\
\noalign{\smallskip}
 ~~SC10--95827~~  & ~~1.974513~~ & ~~2.070361~~ & ~~--0.02345~~ & ~~1.04854~~ & ~~0.076~~ \\
\noalign{\smallskip}
 ~~SC10--132645~~ & ~~1.576676~~ & ~~1.594462~~ & ~~--0.00708~~ & ~~1.01128~~ & ~~0.053~~ \\
\noalign{\smallskip}
 ~~SC10--259946~~ & ~~5.075153~~ & ~~6.183343~~ & ~~--0.03531~~ & ~~1.21836~~ & ~~0.039~~ \\
\noalign{\smallskip}
 ~~SC13--165223~~ & ~~2.043172~~ & ~~2.039852~~ & ~~~~0.00080~~ & ~~0.99838~~ & ~~0.040~~ \\
                  &              & ~~1.230847~~ & ~~~~0.32301~~ & ~~0.60242~~ & ~~0.041~~ \\
\noalign{\smallskip}
 ~~SC13--242700~~ & ~~3.452472~~ & ~~3.262793~~ & ~~~~0.01684~~ & ~~0.94506~~ & ~~0.060~~ \\
                  &              & ~~3.363016~~ & ~~~~0.00771~~ & ~~0.97409~~ & ~~0.077~~ \\
\noalign{\smallskip}
\hline
\end{tabular}
\end{center}

\eject

\begin{center}
{\bf Table 1:} - {\it continued}
\end{center}

\begin{center}
\normalsize
\small
\begin{tabular}{lccccc}
\hline
\noalign{\smallskip}
~~OGLE ID         & $P_1$\th [day]
                                 & $P_{\nu}$\th [day]
                                                & $\Delta f$\th [c/d]
                                                                & $P_{\nu}/P_1$
                                                                              & $A_{\nu}/A_1$
                                                                                          \\
\noalign{\smallskip}
\hline
\noalign{\smallskip}
 ~~SC14--46315~~  & ~~2.315774~~ & ~~1.442102~~ & ~~~~0.26161~~ & ~~0.62273~~ & ~~0.023~~ \\
\noalign{\smallskip}
 ~~SC15--170744~~ & ~~4.992984~~ & ~~3.148984~~ & ~~~~0.11728~~ & ~~0.63068~~ & ~~0.055~~ \\
\noalign{\smallskip}
 ~~SC16--37119~~  & ~~2.795951~~ & ~~3.520911~~ & ~~--0.07364~~ & ~~1.25929~~ & ~~0.051~~ \\
\noalign{\smallskip}
 ~~SC16--177823~~ & ~~1.918044~~ & ~~1.892260~~ & ~~~~0.00710~~ & ~~0.98656~~ & ~~0.046~~ \\
\noalign{\smallskip}
 ~~SC16--194279~~ & ~~2.065767~~ & ~~2.146254~~ & ~~--0.01815~~ & ~~1.03896~~ & ~~0.059~~ \\
\noalign{\smallskip}
 ~~SC16--230207~~ & ~~3.418163~~ & ~~4.416698~~ & ~~--0.06614~~ & ~~1.29213~~ & ~~0.039~~ \\
\noalign{\smallskip}
 ~~SC17--39484~~  & ~~3.541888~~ & ~~4.063954~~ & ~~--0.03627~~ & ~~1.14740~~ & ~~0.055~~ \\
                  &              & ~~4.104648~~ & ~~--0.03871~~ & ~~1.15889~~ & ~~0.045~~ \\
\noalign{\smallskip}
 ~~SC17--39517~~  & ~~2.113535~~ & ~~2.149961~~ & ~~--0.00802~~ & ~~1.01725~~ & ~~0.065~~ \\
                  &              & ~~1.326966~~ & ~~~~0.28046~~ & ~~0.62784~~ & ~~0.046~~ \\
\noalign{\smallskip}
 ~~SC17--80220~~  & ~~2.258854~~ & ~~2.829748~~ & ~~--0.08931~~ & ~~1.25274~~ & ~~0.058~~ \\
\noalign{\smallskip}
 ~~SC17--146711~~ & ~~1.883131~~ & ~~1.137820~~ & ~~~~0.34784~~ & ~~0.60422~~ & ~~0.049~~ \\
\noalign{\smallskip}
 ~~SC17--171481~~ & ~~2.537651~~ & ~~3.707561~~ & ~~--0.12435~~ & ~~1.46102~~ & ~~0.066~~ \\
\noalign{\smallskip}
 ~~SC17--211310~~ & ~~2.023579~~ & ~~2.087774~~ & ~~--0.01519~~ & ~~1.03172~~ & ~~0.054~~ \\
                  &              & ~~2.593589~~ & ~~--0.10861~~ & ~~1.28168~~ & ~~0.037~~ \\
\noalign{\smallskip}
 ~~SC18--144653~~ & ~~4.158193~~ & ~~5.222752~~ & ~~--0.04902~~ & ~~1.25602~~ & ~~0.049~~ \\
\noalign{\smallskip}
 ~~SC18--208875~~ & ~~3.928946~~ & ~~3.508220~~ & ~~~~0.03052~~ & ~~0.89292~~ & ~~0.202~~ \\
\noalign{\smallskip}
 ~~SC19--74265~~  & ~~2.364944~~ & ~~2.790232~~ & ~~--0.06445~~ & ~~1.17983~~ & ~~0.046~~ \\
                  &              & ~~3.080875~~ & ~~--0.09826~~ & ~~1.30273~~ & ~~0.060~~ \\
\noalign{\smallskip}
 ~~SC20--83423~~  & ~~2.068067~~ & ~~2.108720~~ & ~~--0.00932~~ & ~~1.01966~~ & ~~0.053~~ \\
\noalign{\smallskip}
\hline
\end{tabular}
\end{center}

\bigskip\bigskip

\hskip 0pt In most of the FO-$\nu$ Cepheids only one secondary
peak was detected, but in several variables two peaks were found.
In all cases they have extremely small amplitudes. With the
exception of a singe star (SC18--208875), the amplitude ratio
${\rm A}_{\nu}/{\rm A}_{1}$ is always below 0.1, with the average
value of 0.048. We note, that secondary peaks detected in the
first overtone RR~Lyrae stars are typically an order of magnitude
stronger, with ${\rm A}_{\nu}/{\rm A}_{1} = 0.45$ on average
(Alcock et~al. 2000).

\smallskip

\hskip 0pt It is easy to check, that period ratios measured in
FO-$\nu$ Cepheids are not compatible with those of the radial
modes. This implies, that the secondary frequencies detected in
these pulsators must correspond to {\bf nonradial modes of
oscillations}.

\smallskip

\hskip 0pt The secondary frequencies in FO-$\nu$ Cepheids come in
two different flavours. In 37 variables they are located close to
the primary pulsation frequency ${\rm f}_{1}$. Several examples of
such behaviour are displayed in Fig.\th 1. In 84\% of cases,
secondary frequencies are lower than the primary one ($\Delta{\rm
f} < 0$). When two secondary peaks are detected, they always
appear on the same side of the primary peak. In 7 FO Cepheids a
different type of secondary periodicity was found: a {\it high
frequency} mode, with the period ratio of ${\rm P}_{\nu}/{\rm
P}_{1} = 0.60 - 0.64$. The two types of secondary nonradial modes
are not mutually exclusive. In two objects (SC13--165223 and
SC17--39517) both a high frequency secondary peak and a secondary
peak close to the primary frequency are present.

\vskip 28truecm

\includegraphics{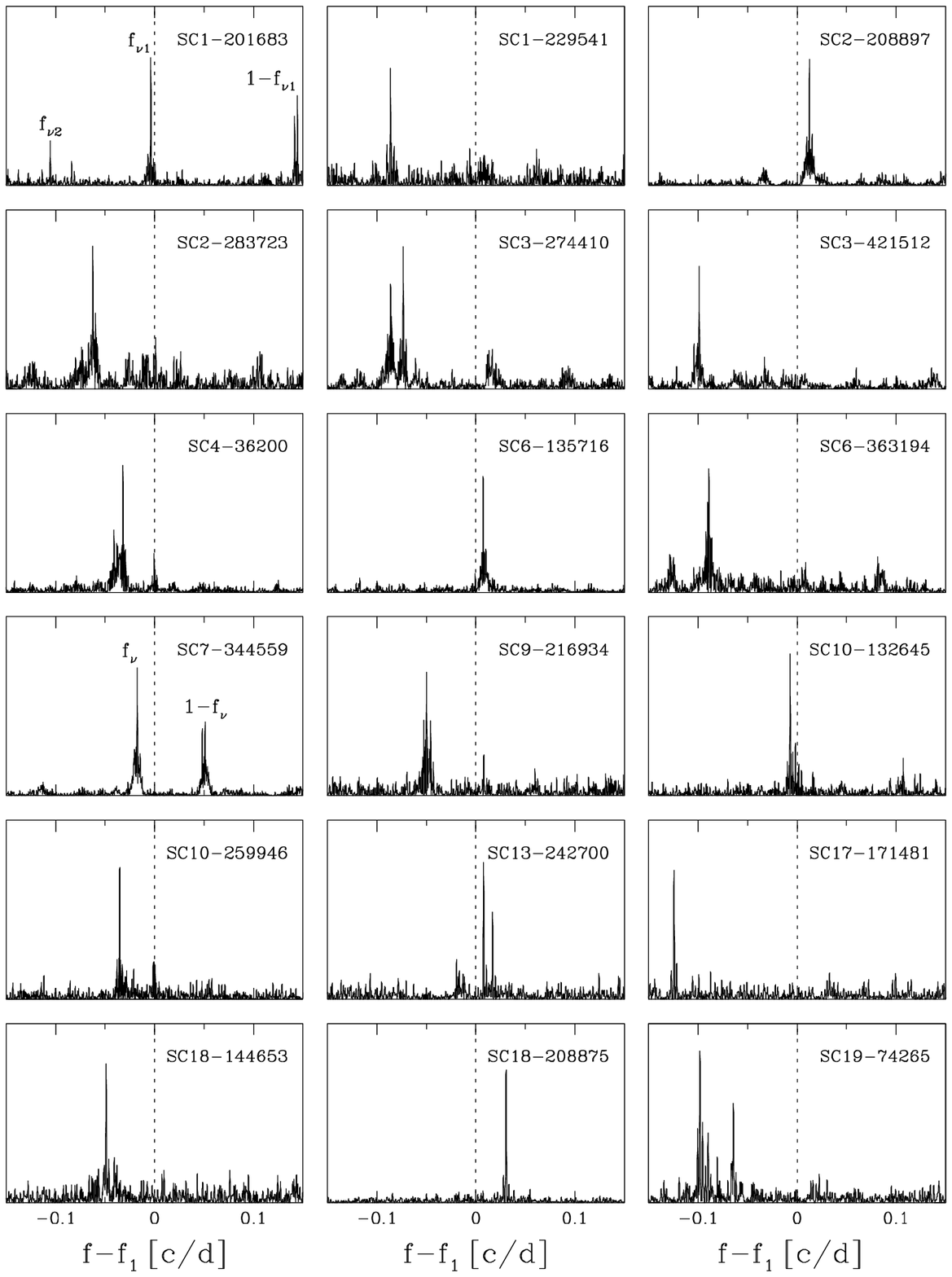}
\begin{figure}[h]
\vskip 16truecm
\caption{Prewhitened power spectra of FO-$\nu$ Cepheids.
Frequencies of removed radial modes indicated by dashed lines.}
\end{figure}

\bigskip

\hskip 0pt In Fig.\th 2 we show the distribution of frequency
differences, $\Delta{\rm f}$, for the LMC FO-$\nu$ Cepheids and,
for comparison, for the LMC first overtone RR~Lyrae stars (Alcock
et~al. 2000). The two distributions are not very different.
FO-$\nu$ Cepheids show somewhat stronger preference for negative
$\Delta{\rm f}$, but other\-wize, in both types of overtone
pulsators nonradial modes are found in similar distances from the
radial mode. The only difference between the two histograms is the
presence of high frequency secondary peaks (${\rm P}_{\nu}/{\rm
P}_{1}\sim 0.62$) in the Cepheids, but not in the RR~Lyrae stars.

\hskip 0pt Although the population of FO Cepheids in the LMC
extends down to periods as short as 0.4\th day, nonradial modes
were detected only in stars with ${\rm P}_{1} > 1.2$\th day. In
fact, the incidence rate of nonradial modes systematically
increases with the primary pulsation period, reaching 19\% for
stars with ${\rm P}_{1} > 3.0$\th day. This is illustrated in
Fig.\th 3. We interpret this behaviour as a selection effect: the
Cepheids with longer periods are brighter, consequently it is
easier to detect very low amplitude secondary periodicities in
their lightcurves. If so, then the true incidence rate of
nonradial modes in LMC overtone Cepheids can be significantly
higher that 9\% derived here for the entire OGLE-II sample.

\bigskip

\vskip 12truecm

\includegraphics{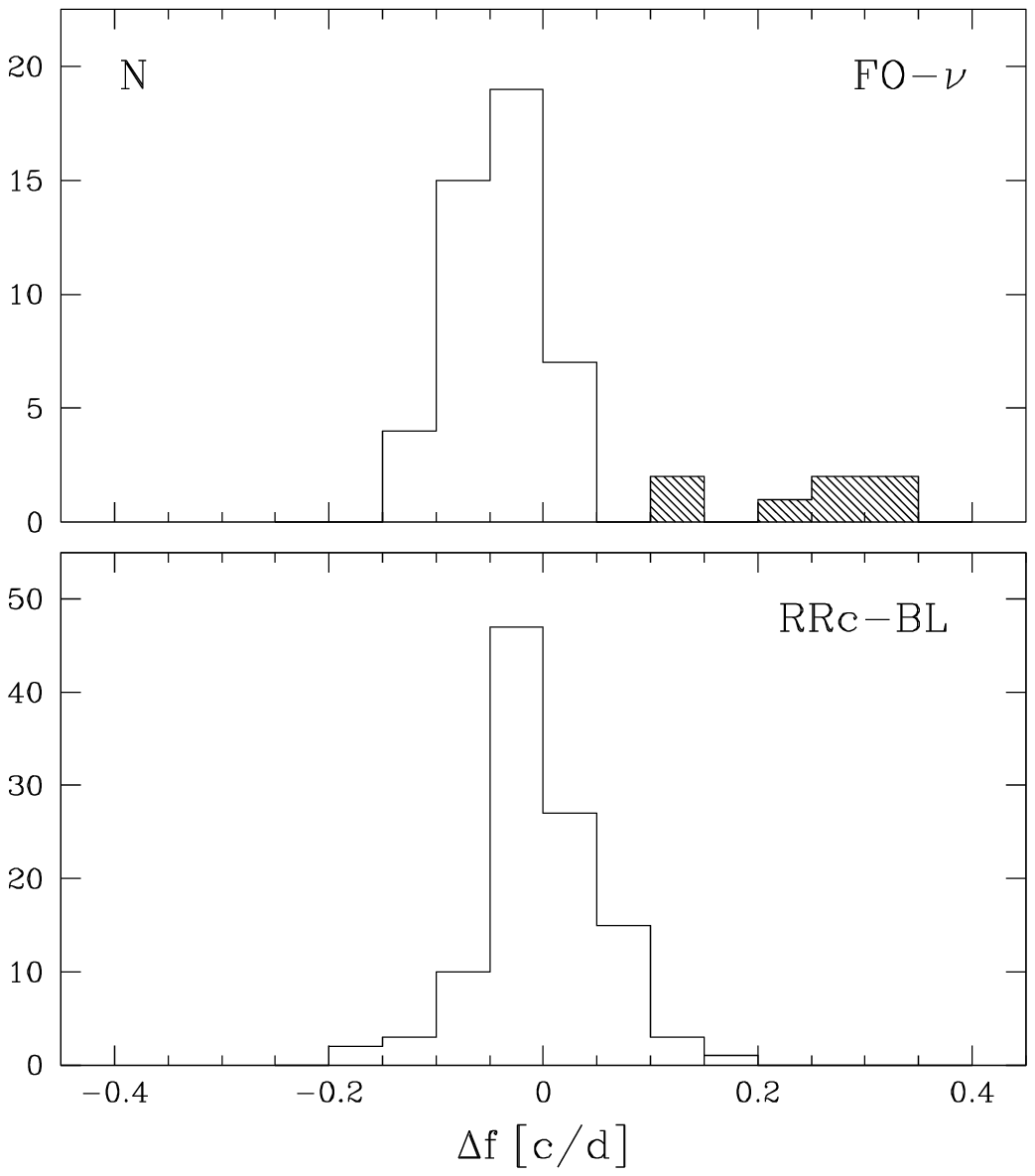}
\begin{figure}[h]
\vskip -2.9truecm
\caption{Distribution of frequency differences $\Delta f = f_{\nu} - f_{1}$ for
LMC FO-$\nu$ Cepheids and RRc-BL stars. High frequency nonradial modes shaded.}
\end{figure}

\bigskip

\vskip 10truecm

\includegraphics{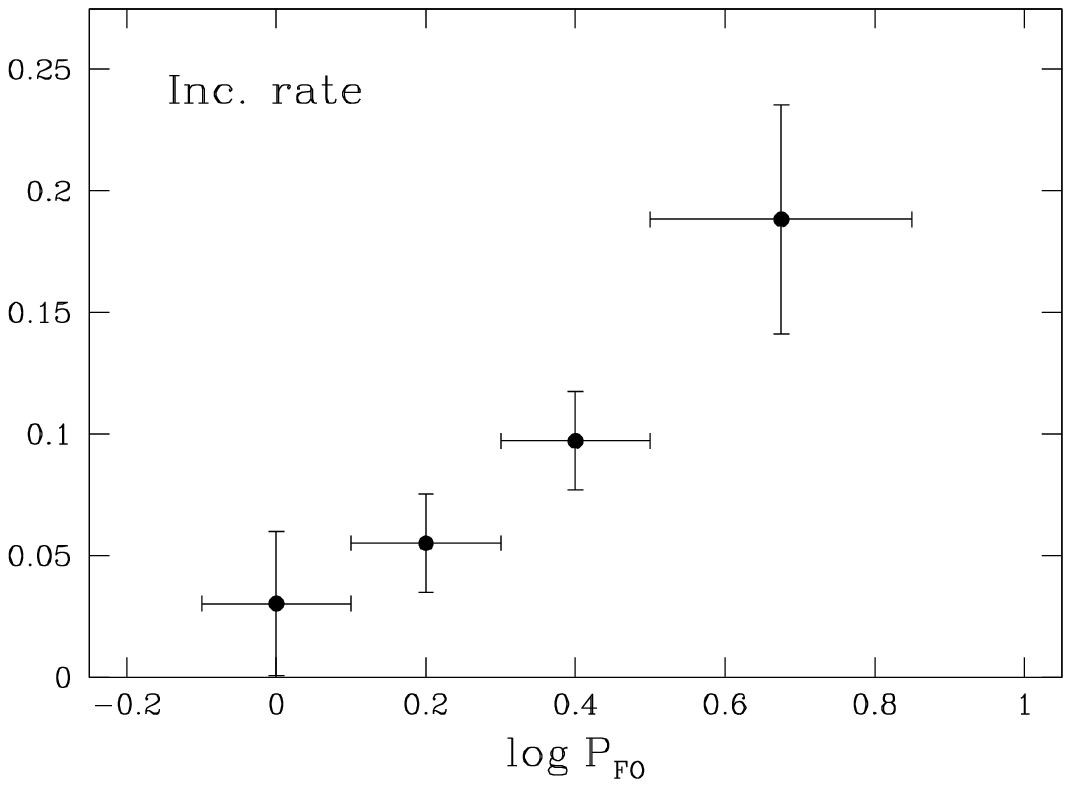}
\begin{figure}[h]
\vskip -4.0truecm
\caption{Incidence rate of LMC FO-$\nu$ Cepheids {\it vs.} primary
period.}
\end{figure}

\bigskip

\section*{Fundamental Mode Cepheids}

\hskip 0pt OGLE-II catalogs list 719 fundamental mode (FU)
Cepheids. We have searched all of them for secondary
periodicities. We have found {\it no nonradial modes} in the FU
Cepheids of the LMC.

\section*{FU/FO Double-Mode Cepheids}

\hskip 0pt In the course of systematic analysis of OGLE-II
Cepheids, we have discovered 4 new FU/FO double mode pulsators.
Together with stars listed in OGLE-II catalogs (Soszy\'nski et~al.
2000) this brings to 23 the total number of known FU/FO Cepheids
in the LMC. We have found nonradial modes in 3 of them. These are
the first detections of nonradial modes in the FU/FO double-mode
Cepheids. In the following, we call these stars FU/FO-$\nu$
Cepheids. The prewhitened power spectra of the 3 stars are
displayed in Fig.\th 4. In the first two Cepheids, the secondary
mode appears very close to the first overtone radial mode. The
values of the frequency differences $\Delta{\rm f} = {\rm f}_{\nu}
- {\rm f}_{1} $ are very similar to those observed in the FO-$\nu$
Cepheids. In the third star, the secondary mode has been found at
high frequency, with the period ratio of ${\rm P}_{\nu}/{\rm
P}_{\rm FO} = 0.623$, {\it i.e.} with the same strange ratio which
is frequently observed in the FO-$\nu$ Cepheids. Clearly,
nonradial modes excited in the FU/FO-$\nu$ Cepheids are somehow
connected with the first radial overtone. Their frequencies are
drawn from the same distribution as in the single-mode FO-$\nu$
Cepheids.

\bigskip

\vskip 10truecm

\includegraphics{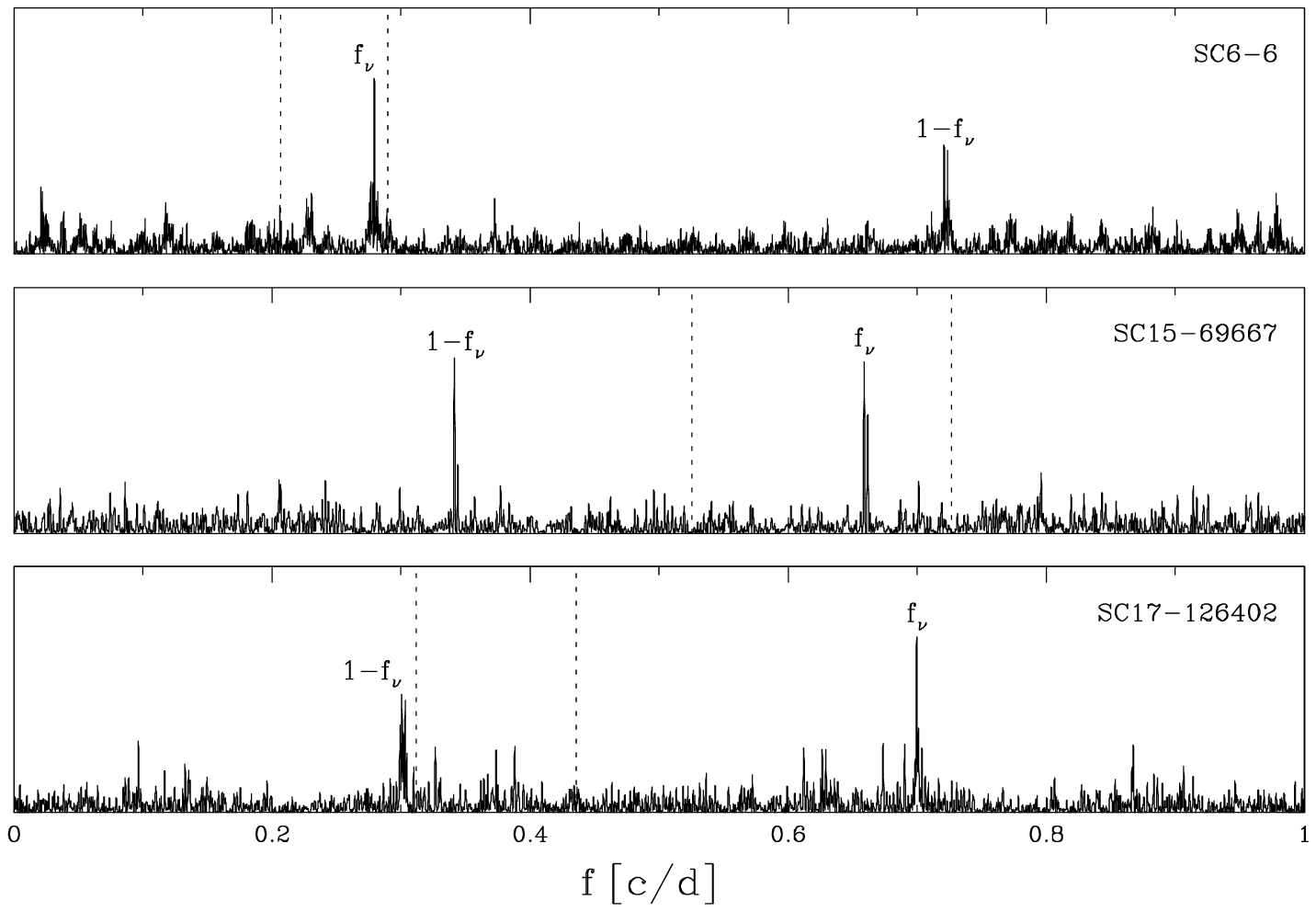}
\begin{figure}[h]
\vskip -1truecm
\caption{Prewhitened power spectra of FU/FO-$\nu$ Cepheids.
Frequencies of removed radial modes indicated by dashed lines.}
\end{figure}

\bigskip

\section*{Comparison with RR~Lyrae Stars}

\hskip 0pt In the last decade nonradial modes have been detected
in many RR~Lyrae stars belonging to various stellar systems,
including both Magellanic Clouds and the Galactic Bulge. Since
pulsations of Cepheids and of RR~Lyrae stars are in many ways
similar, it is instructive to compare the properties of nonradial
modes in these two types of stars. Several obvious differences
should be pointed out:

\begin{itemize}
\item Amplitudes of nonradial modes in classical Cepheids are an order of
                 magnitude {\it smaller} than in RR~Lyrae stars.
\item In RR~Lyrae stars, nonradial modes are detected both in fundamental mode
                 pulsators (RRab) and in first overtone pulsators (RRc). In
                 classical Cepheids nonradial modes are detected {\it only} in
                 first overtone pulsators.
\item Nonradial modes are detected in 3 FU/FO double-mode Cepheids, which
                 constitutes $\sim 13$\% of the LMC sample. This type of
                 pulsators are extremely rare among RR~Lyrae stars. In more
                 than 200 FU/FO RR~Lyrae variables known in various steller
                 systems, {\it only one} detection of nonradial mode has been
                 reported (Alcock et~al. 2000).
\item When two secondary frequencies are found in an RR~Lyrae star, they
                 usually form, together with the primary frequency, an equally
                 spaced triplet. In sharp contrast, equidistant triplets are
                 {\it never} observed in classical Cepheids.
\item Nonradial modes in Cepheids are usually detected very close to the primary
                 pulsation mode, but in several stars a high frequency mode with
                 ${\rm P}_{\nu}/{\rm P}_{1}\sim 0.62$ was found. Such high
                 frequency modes are {\it not observed} in RR~Lyrae pulsators.
\end{itemize}

\bigskip

\section*{References}

{\everypar={\hangindent=1.5truecm}
\parindent=0pt\frenchspacing

Alcock,~C., Allsman,~R.~A., Alves,~D.~R., et~al. 2000, ApJ, 542,~257

Allsman,~R.~A. \& Axelrod,~T.~S. 2001, astro-ph/0108444

Moskalik,~P. \& Ko{\l}aczkowski,~Z. 2008, in M.-J.~Goupil ed., Nonlinear
     stellar hydrodynamics, EAS Publication Series (in~press)

Moskalik,~P., Ko{\l}aczkowski,~Z. \& Mizerski,~T. 2004, in D.~W.~Kurtz \&
     K.~Pollard eds., Variable Stars in the Local Group, IAU~Coll.~193,
     ASP~Conf. Ser., 310,~498

Moskalik,~P., Ko{\l}aczkowski,~Z. \& Mizerski,~T. 2006, in A.~R.~Walker \&
     G.~Bono eds., Stellar Pulsation and Evolution,
     Mem.~Soc.~Astron.~Ital., 77,~563

Soszy\'nski,~I., Udalski,~A., Szyma\'nski,~M., et~al. 2000,
     Acta~Astron., 50,~451

Udalski,~A., Soszy\'nski,~I., Szyma\'nski,~M., et~al. 1999,
     Acta~Astron., 49,~223

\.Zebru\'n,~K., Soszy\'nski,~I., Wo\'zniak,~P.~R. et~al. 2001,
     Acta~Astron., 51,~317
}

\vspace{1.0cm}

\begin{flushleft}
{\bf Acknowledgments.} This work has been supported by the Polish
MNiSW Grant No. 1 P03D 011 30.
\end{flushleft}

\end{document}